\RequirePackage{lineno}
\documentclass[aps,prl,twocolumn,amsmath,amssymb]{revtex4}

\usepackage{graphicx}
\usepackage{lineno}
\usepackage[colorlinks,linkcolor=blue,anchorcolor=blue,citecolor=blue]{hyperref}
\begin{document}


\title{Superconductivity Induced at a Point Contact on the Topological Semimetal Tungsten Carbide}

\author{Xing-yuan Hou$^{1,2}$, Zong Wang$^{2,3}$,Ya-dong Gu$^{2,3}$}
\author{Jun-bao He$^2$,Dong Chen$^2$,Wen-liang Zhu$^{2,3}$, Meng-di Zhang$^{2,3}$, Fan Zhang$^{2,3}$, Yuan-feng Xu$^{2,3}$, Shuai Zhang$^{2}$, Huai-xin Yang$^{2}$, Zhi-an Ren$^{2,3,5}$}
\author{Hong-ming Weng$^{2,5}$}
\author{Ning Hao$^{6}$}
\author{Wen-gang Lv$^{2}$}\email{wglu@iphy.ac.cn}
\author{Jiang-ping Hu$^{2,3,5}$}
\author{Gen-fu Chen$^{2,3,5}$}\email{gfchen@iphy.ac.cn}
\author{Lei Shan$^{1,2,3,4}$}\email{lshan@ahu.edu.cn}

\affiliation{$^1$Institutes of Physical Science and Information Technology, Anhui University, Hefei 230601, China}

\affiliation{$^2$Beijing National Laboratory for Condensed Matter Physics, Institute of Physics, Chinese Academy of Sciences, Beijing 100190, China}

\affiliation{$^3$School of Physical Sciences, University of Chinese Academy of Sciences, Beijing 100190, China}

\affiliation{$^4$Key Laboratory of Structure and Functional Regulation of Hybrid Materials(Anhui University),Ministry of Education,Hefei,230601,P.R.China}

\affiliation{$^5$Songshan Lake Materials Laboratory, Dongguan, Guangdong 523808, China}

\affiliation{$^6$Anhui Province Key Laboratory of Condensed Matter Physics at Extreme Conditions, High Magnetic Field Laboratory, Chinese Academy of Sciences, Hefei 230031, China}

\date{\today}

\begin{abstract}

We report the observation of local superconductivity induced at the point contact formed between a normal metal tip and WC -- a triple point topological semimetal with super hardness. Remarkably, the maximum critical temperature is up to near 12 K but insensitive to the tip's magnetism. The lateral dimensions of the superconducting puddles were evaluated and the temperature dependencies of superconducting gap and upper critical field were also obtained. These results put constraints on the explanation of the induced superconductivity and pave a pathway for exploring topological superconductivity.

\end{abstract}


\maketitle

\setpagewiselinenumbers 
\modulolinenumbers[1]  


Unconventional superconductors and topological materials have attracted tremendous attentions due to their novel physical properties and potential applications. Integration of superconductivity and topological nature may lead to new emergent phenomena such as exotic topological superconductivity, for the interesting interplay between electron pairing and the spin-orbit coupling \cite{10TI_ZhangSC,11Majorana_Leijnse,12Majorana_Alicea,13TI_Ando}. Recently, superconductivity was induced at a point contact between a normal metal tip and the 3D Dirac semimetal Cd$_3$As$_2$ \cite{1Cd3As2_GS,2Cd3As2_WJ}. Soon after, similar phenomenon was reproduced in the Weyl semimetal TaAs \cite{3TaAs_GS,4TaAs_WJ}. In particular, the induced superconductivity could coexist with preserved topological properties in the line-nodal semimetal ZrSiS \cite{TISCinZrSiS_GSheet_2018}. Such tip-induced superconductivity (TISC) is very exciting since it is extremely rare in common metals and provides a unique way of searching topological superconductivity \cite{review_JWang_TISC_2018}. Although the most likely origins of the TISC were speculated to be tip pressure-induced band reconstruction, confinement effect, charge carrier doping \cite{3TaAs_GS}, or remarkable enhancement of density of states under the tip \cite{TISCinZrSiS_GSheet_2018}, the dominant mechanism remains elusive. Meanwhile, unconventional superconductivity has been proposed for Cd$_3$As$_2$, yet based on two distinct experimental observations, i.e., a pseudogap behavior above the critical temperature \cite{1Cd3As2_GS} or an abnormal spectral shape at low temperatures \cite{2Cd3As2_WJ}. For TaAs, such unconventionality is still in debate \cite{3TaAs_GS, 4TaAs_WJ}. Therefore, more evidences are desired to clarify this issue.

Most recently, a new type of topological quasiparticles was proposed for the metals hosting symmetry-protected three-band crossings (three-fold degeneracy) near the Fermi level \cite{5Triple_point_ZhuZM,6Triple_point_WengHM1,6Triple_point_WengHM2}. These triple point fermions were first observed in MoP \cite{7MoP_DingHong} and the non-trivial topology of the new semimetal state was identified in WC subsequently \cite{8WC_DingHong}. As a triple point topological semimetal, WC can be regarded as an intermediate state between Dirac and Weyl semimetals with four- and two-fold degenerate points, respectively. Furthermore, the super hardness of WC ensures its topological band structure to be robust against tip pressure, providing an ideal platform to inspect the universality and mechanism of TISC in topological semimetals.

In this work, we performed point contact experiments on WC using both non-magnetic and ferromagnetic normal metal tips. Local superconductivity with various lateral sizes could be induced for all these tips, providing a strong evidence of its unconventional nature. In addition, temperature dependent superconducting gap $\Delta(T)$ and upper critical field $H_c(T)$ were obtained. These observations are helpful to reveal the mechanism of TISC in topological semimetals and deserve to be studied intensively.

\begin{figure}[]
\hspace{0cm}\includegraphics[scale=0.45]{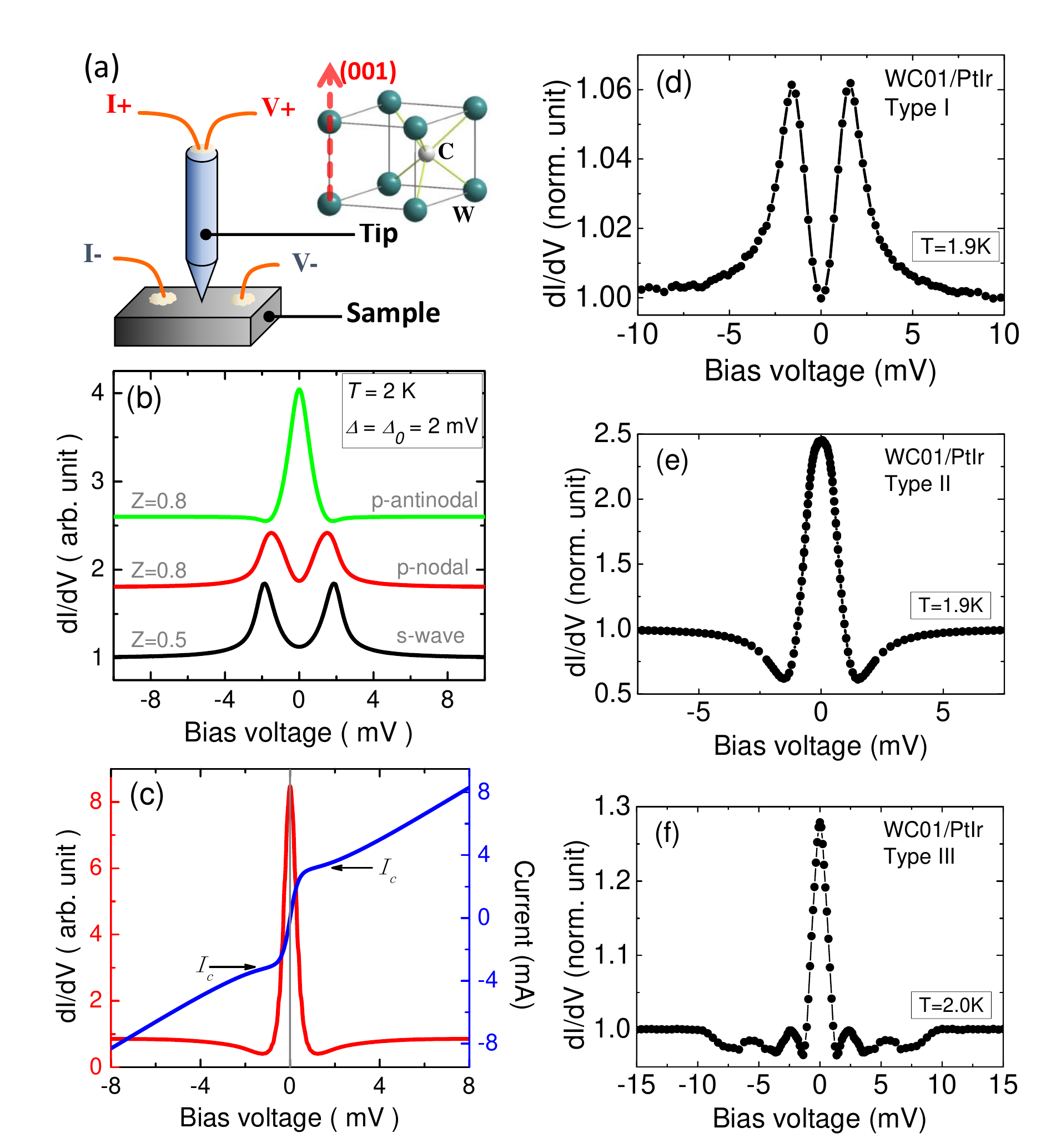}
\caption{\label{fig:TISC} (Color online) (a) Schematic diagram of the PC spectra measurement configuration. The current injection is oriented along the [001] axis indicated in the crystal structure of WC. (b) Some typical point contact spectra calculated for an s-wave (black) and a p-wave superconductor with current injections along nodal (red) and anti-nodal (green) directions, where $T$, $Z$, $\Delta$ and $\Delta_0$ are temperature, effective interfacial barrier height, s-wave superconducting gap and maximum gap in the p-wave form of $\Delta_{\theta}=\Delta_{0}sin(\theta)$.  (c) Current vs. voltage characteristics of a PC lying in the thermal regime and the $dI/dV$ spectrum obtained by numerical differentiation.  (d)-(f) Typical PC spectra obtained with PtIr tips, labeled as Type I, II and III, respectively (see Text).}
\end{figure}

Point contact measurements were carried out in our home-built probes based on a physical property measurement system (PPMS) from Quantum Design. We adopted here the most common point contact (PC) configuration, often called ``needle-anvil" as illustrated in Fig.~\ref{fig:TISC}(a). An ultrafine differential micrometer (Siskiyou 100tpi-05d) or an Attocube nanopositioner stack was used to control tip approaching towards a sample along the longitudinal axis of PPMS, and a pair of bevel gears was used for tip approaching horizontally. Point contacts (PCs) between various metal tips and three crystalline WC specimens (labeled as WC01, WC02 and WC03) were formed in situ with current injecting along the [001]-axis of WC crystals. During the spectral measurements, current-voltage ($I \sim V$) curves and differential conductance spectra ($ dI/dV \sim V$) were recorded simultaneously.


\begin{figure}[]
\hspace{0cm}\includegraphics[scale=0.45]{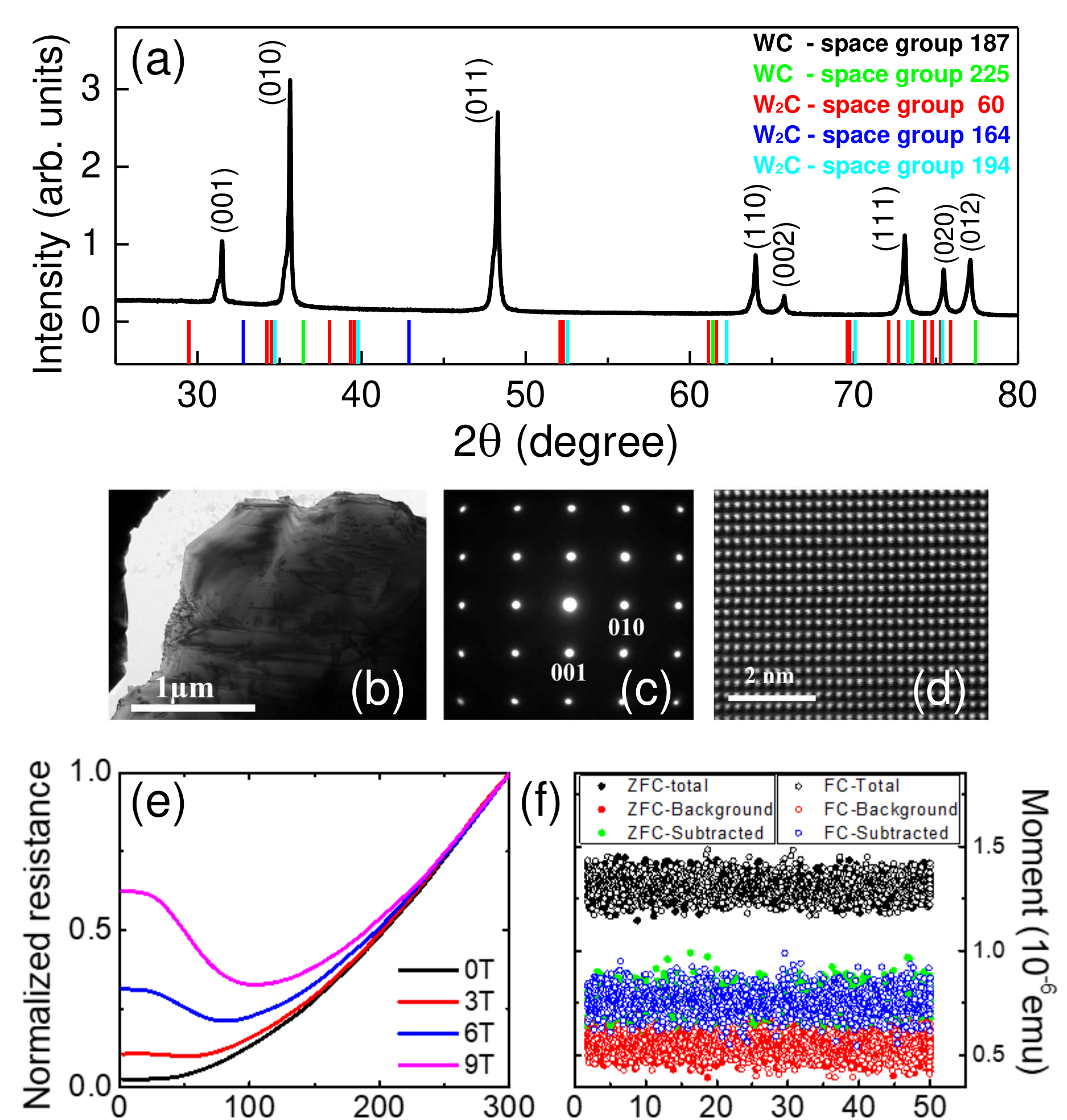}
\caption{\label{fig:sample} (Color online) (a) Powder XRD pattern of WC obtained by grinding lots of single crystals at room temperature. (b) Bright field TEM image taken along the [100] zone axis directions for the WC single crystal. (c) and (d) Electron diffraction pattern and the corresponding HAADF STEM image taken along the [100] zone axis directions for the WC single crystal. (e) Temperature dependent resistance of WC measured in different magnetic fields. (f) Raw data of magnetization versus temperature down to 2K measured for WC in zero field cooling (ZFC) and field cooling (FC) processes respectively.}
\end{figure}

The high-quality single crystals of WC studied here were grown by the flux method \cite{23WC_MR}, with a dimension around 2 mm$^2$ and a thicknesses from 0.3 mm to 0.65 mm. The high purity of the WC samples has been confirmed by XRD and TEM measurements. Fig.\ref{fig:sample}(a) shows the room-temperature powder XRD pattern of WC obtained by grinding lots of single crystals synthesized in this work. All the diffraction peaks can be well indexed by the hexagonal structure of WC with a space group of $P\overline6m2$ (No.187). We didn't find any peaks indexed by the other structures of WC or W$_2$C with space groups shown in Fig.\ref{fig:sample}(a) \cite{1, 2, 3}. Figures \ref{fig:sample}(b) - (d) shows the typical bright field TEM image, electron diffraction pattern and the corresponding HAADF STEM image taken along the [100] zone axis direction. The repeated TEM measurements reveal no signatures of any impurity phase rather than the expected hexagonal WC. As shown in Fig. \ref{fig:sample}(e), the temperature dependent resistance curves in various magnetic fields show large magnetoresistance at low temperature, consistent with the previous measurement\cite{23WC_MR}, whereas no superconducting transition has been detected down to the lowest temperature of 2 K. Figure \ref{fig:sample}(f) displays the raw data of magnetization versus temperature, in which both total signal and background signal for empty holder are provided. Obtained by subtracting background signals from total signals, the nominal sample signals are of the order of 10$^{-7}$ emu, around the minimal detectivity of the Quantum Design MPMS system. In the narrow temperature range studied here, the nominal sample signals are temperature independent, indicating again no superconductivity existing in the sample.

In general, $I \sim V$ curves of a PC between two normal metals (N/N) look more like straight lines at low bias voltages due to its nearly constant resistance, and no obvious features can be seen in the PC spectra $dI/dV \sim V$. The situation is very different for a normal metal/superconductor (N/S) PC. When superconducting state is turned on, the spectral shape will depend on the superconducting pairing symmetry and the effective barrier height at the N/S interface, which could be described by a generalized Blonder-Tinkham-Klapwijk (BTK) formalism \cite{15BTK, 16Dynes, 17BTK_Gamma, 18d-BTK, 19PCS_Cuprate}. A PC spectrum with lower barrier height $Z$ is usually referred to as Andreev reflection spectrum (the bottom line in Fig.~\ref{fig:TISC}(b)). Since Andreev reflection roots in the formation of Cooper pairs in a superconductor, it is a unique and convincing evidence of superconductivity\cite{review_MultiBand, 4}. Although the p-wave spectrum with current injection along nodal direction is similar to the curve of s-wave (the middle line in Fig. \ref{fig:TISC}(b)), the anti-nodal spectrum shows a significant zero-bias conductance peak (ZBCP) due to the formation of Andreev bound states (the top line in Fig.\ref{fig:TISC}(b)), which have usually been utilized to identify unconventional superconductivity\cite{review_MultiBand, 19PCS_Cuprate}. However, it should be pointed out that, the above-mentioned cases are based on the ideal ballistic limit, i.e., the contact radius is much smaller than the electron mean free path ($a\ll l$).  If $a\gg l$, i.e., the thermal regime is dominant, a non-intrinsic ZBCP could often be observed for both nodeless and nodal superconductors. In this regime, the inelastic scattering of electrons in the contact region will cause Joule heating and a local rise in temperature with increasing current \cite{PCS_review_Jansen1980, review_PCS_Duif1989, review_MultiBand}. A superconducting to normal transition will occur at a specific current and a spurious ZBCP arises in the PC spectrum, which is called ``critical current effect" \cite{Critical_Westbrook_1999, 20MgCNi3, 21Ic_GS, ZBCP_Gifford_2016}, as illustrated in Fig.~\ref{fig:TISC}(c). From this perspective, a measured ZBCP cannot be a conclusive evidence of unconventional superconductivity.



\begin{figure}[]
\includegraphics[scale=0.48]{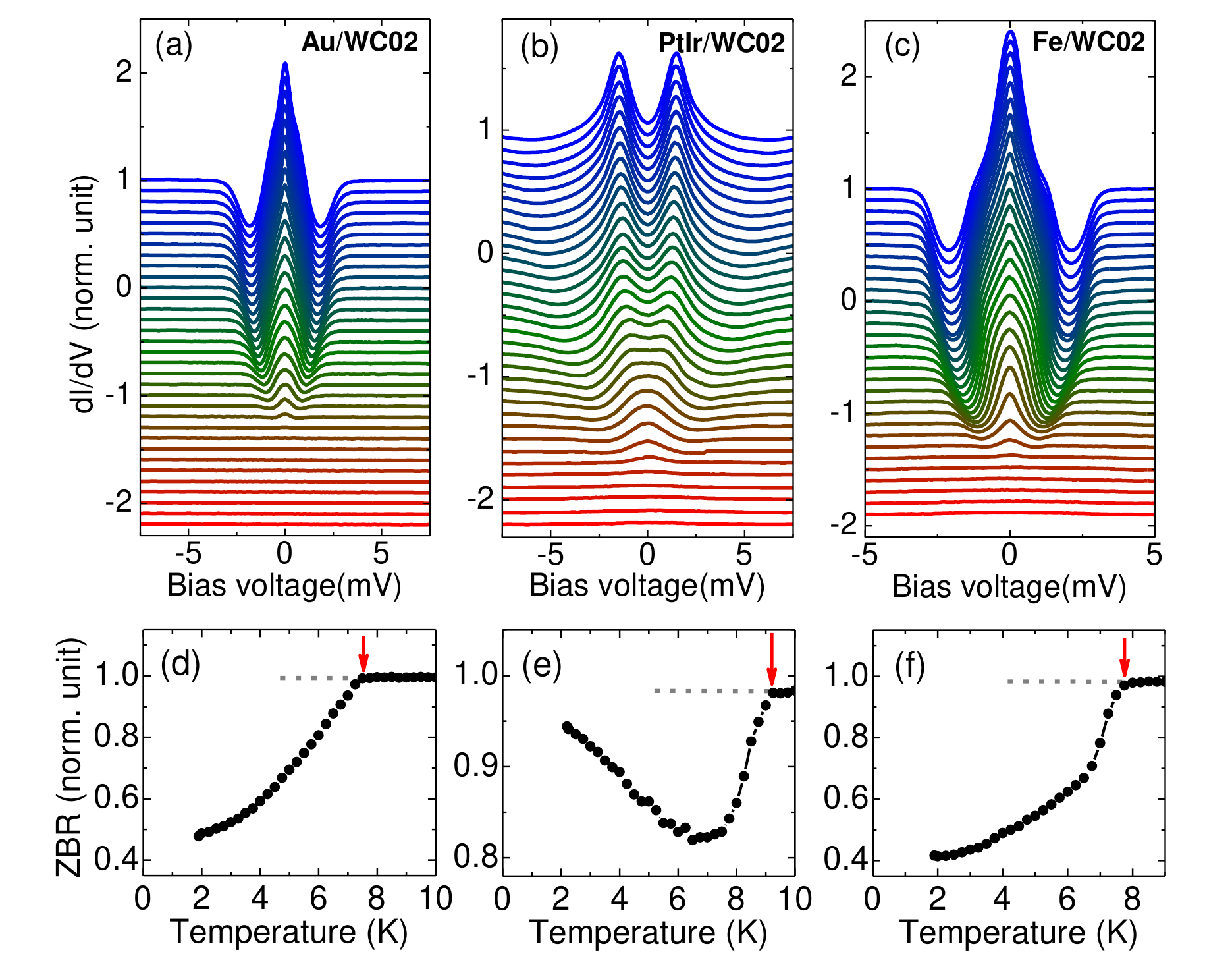}
\caption{\label{fig:VT} (Color online) (a)-(c), Temperature evolutions of spectra measured at different PCs with various tips of Au, PtIr alloy and Fe. (d)-(f) Zero-bias resistance (ZBR) taken from the data in (a)-(c), respectively.}
\end{figure}

\begin{figure}[]
\includegraphics[scale=0.45]{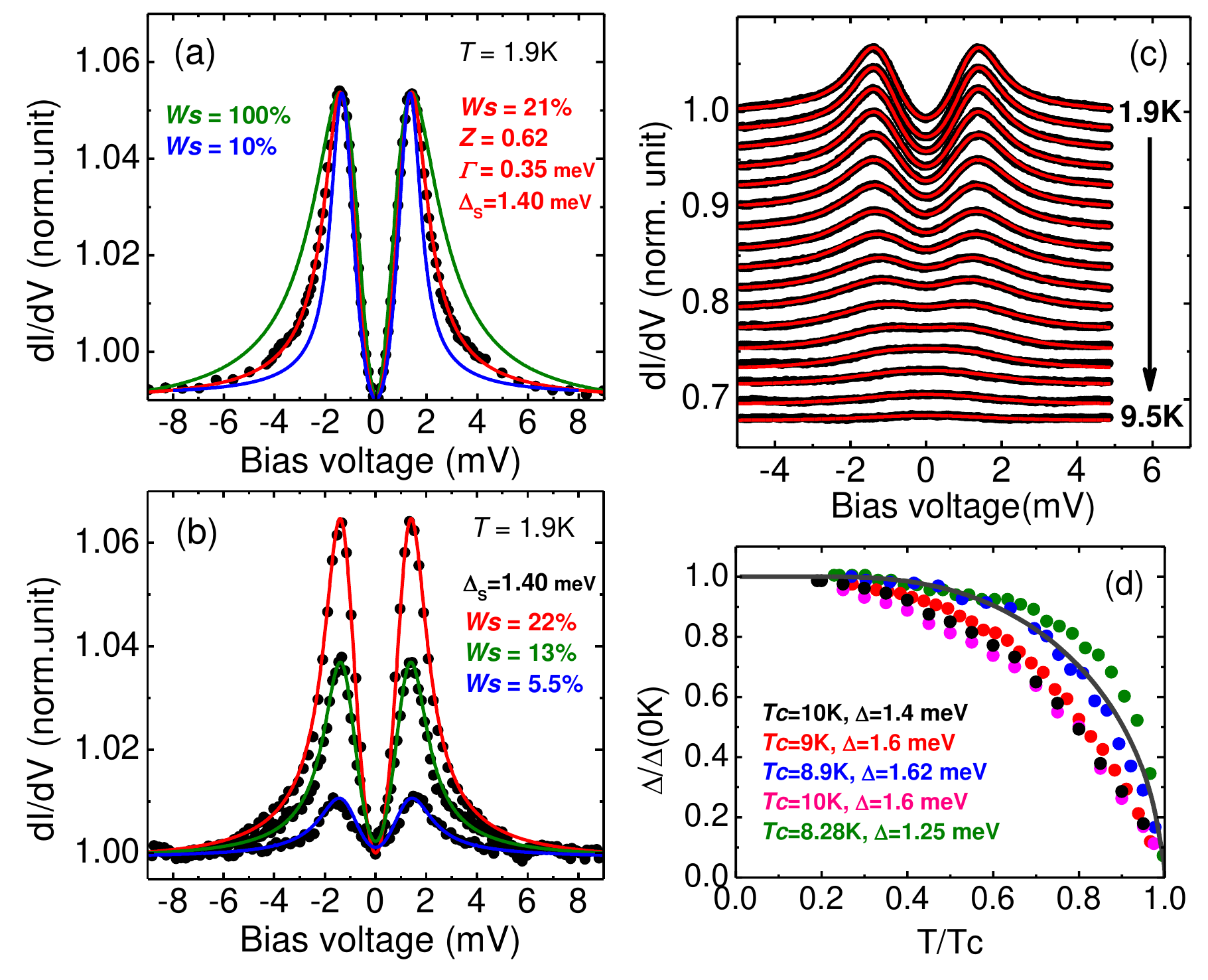}
\caption{\label{fig:BTK} (Color online) Fitting the measured PC spectra to the revised BTK model taking into account the normal-state channel. (a) Comparison of the fitting results with different weight of the superconducting channel ($W_S$). (b) Spectra obtained by fine-tuning junction resistance at the same location and the calculated curves using different $W_S$ values. (c) A global fitting of a series of temperature dependent PC spectra using the revised BTK model with the parameters of $Z=0.64\pm0.02$, $\Gamma=0.3\pm0.02$ and $W_S=22\%$ for all the temperatures. (d) Temperature dependencies of superconducting gaps obtained for various point contacts. }
\end{figure}

Though the as-grown WC single crystal is not superconductive, it was found that superconductivity could be induced at the point contact formed between a normal metal tip and WC with a good repeatability involving all the used tips and the studied samples. As demonstrated in Figs.~\ref{fig:TISC}(d)-(f), the obtained spectra can be classified into three categories: (I) with a typical Andreev reflection spectral shape, (II) showing a zero-bias conductance peak (ZBCP) and two symmetric dips beside it, and (III) having a ZBCP with multiple pairs of dips.
As above-mentioned, unconventional pairing symmetries could give a ZBCP by generating zero-energy bound states at N/S interface \cite{19PCS_Cuprate}, or alternatively, topological superconductors could contribute significant zero-energy density of states at boundaries or impurities due to their nontrivial topology \cite{10TI_ZhangSC, 11Majorana_Leijnse, 12Majorana_Alicea, 13TI_Ando}. For the latter two cases, the symmetric dips beside the ZBCP locate near the superconducting gap energy \cite{18d-BTK, 24TI_tunneling}, thus don't depend on the junction resistance if only the gap is unchangeable. We find that some of the experimental data coincide with this expectation, but some other results exhibit mutable dip energy by changing junction resistance. Although an additional resistance inherent in pseudo-four-probe electrical measurements could lead to a spurious uncertainty of the characteristic energies \cite{25additionalR}, we believe that critical current effect should be involved in, at least, some of our measurements. Based on these knowledge, it could be deduced that the spectra of type III are contributed by multiple parallel channels, corresponding to different micro-constrictions in the large contact area \cite{26MgB2_2gap, 27PCS_magnetic, 28PCS_Vortex}. This mechanism can be further verified by the data in Figure \ref{fig:BTK}.

\begin{figure}[]
\includegraphics[scale=0.45]{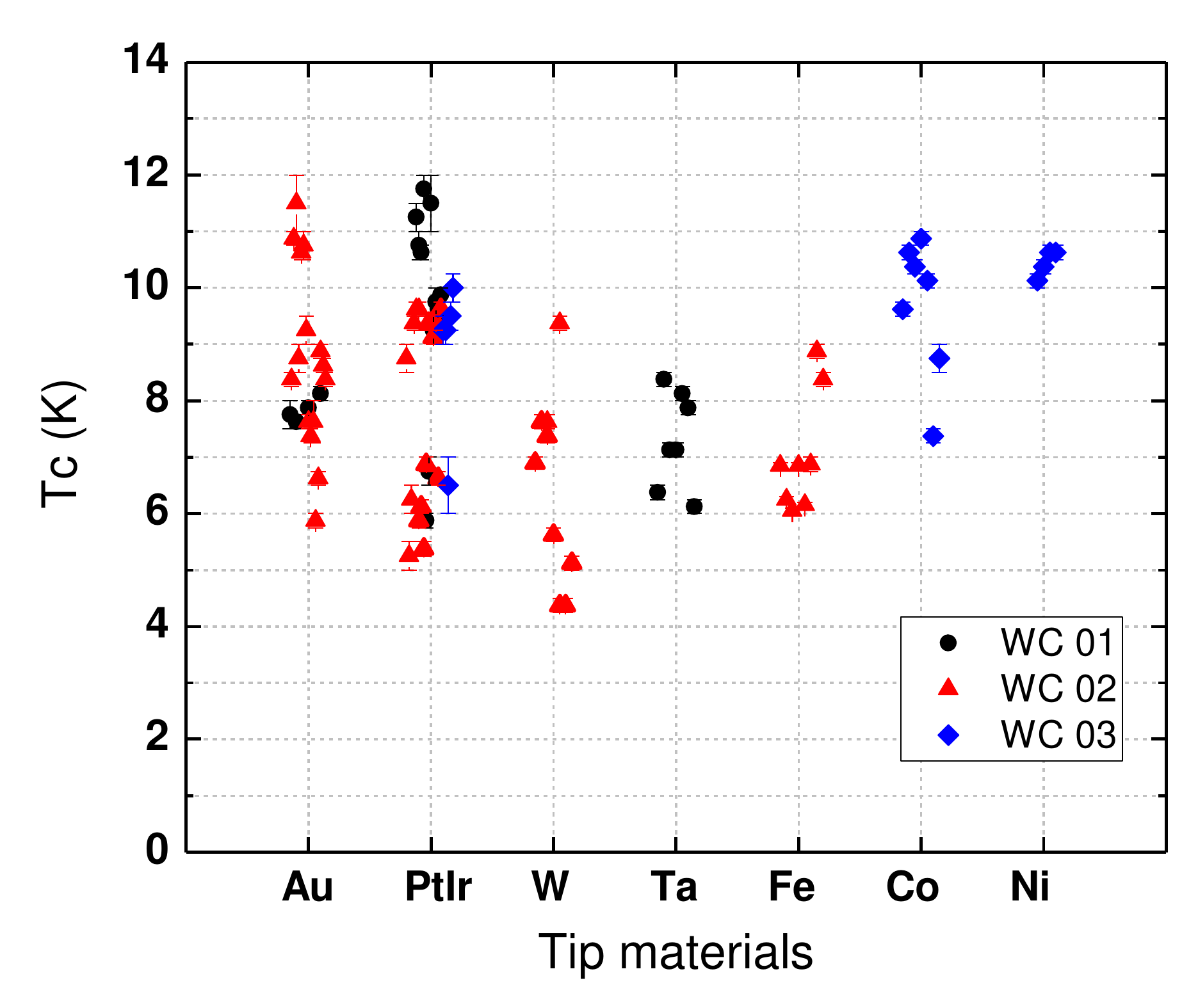}
\caption{\label{fig:Statistics} (Color online) Critical temperatures of TISC realized on WC with diverse metal tips. Statistics were performed on all three WC crystals. Each critical temperature was determined by the vanishing point of superconducting signal in a set of temperature dependent spectra. }
\end{figure}

Figures \ref{fig:VT}(a)-(c) show temperature dependent spectra for three PCs prepared with the non-magnetic Au and PtIr tips and a ferromagnetic Fe tip, respectively. Both the Andreev reflection and ZBCP signals vanish at the corresponding superconducting critical temperature, therefore $T_c$ can be determined by the onset of the resistive transition. In order to highlight superconducting transition and determine $T_c$, the zero-bias resistance (ZBR) derived from Figs.~\ref{fig:VT}(a)-(c) is plotted against temperature in Figs.~\ref{fig:VT}(d)-(f). It can be seen that for multiple-channel PCs, more than one transitions might be observed. All $T_c$ values obtained in this work are summarized in Fig.~\ref{fig:Statistics}, with the maximum value close to 12 K and the minimum one about 4 K. The wide distribution of critical temperatures especially for the different channels of a specified PC indicates that the induced superconductivity is quite local and depends on the real condition of each microcontact in the contact area. Such local superconductivity is difficult to be detected by macroscopic measurements of resistance and magnetization, hence additional techniques other than point contact are necessary to rebuild it in a larger scale. As a preliminary attempt, we have made some progress in metal film deposition on WC crystals \cite{Hetero_Zhu_2018}. However, a connected superconducting area large enough is yet to be achieved.


Another interesting finding here is that local superconductivity can be triggered on non-superconducting WC by ferromagnetic Fe, Co and Ni tips as well as the nonmagnetic tips made of PtIr alloy, Au, W, and Ta, providing a strong evidence for its unconventional nature. The measured signals are significant with a good repeatability, ruling out experimental occasionality. Remarkably, the maximal $T_c$ achieved by a particular metal tip is not sensitive (almost irrelevant) to the tips' magnetism. This is surprising because it is well known that ferromagnetism will break spin singlet pairing and suppress conventional superconductivity, especially for a thin film similar to the interfacial superconductivity observed here. According to the previous works on conventional superconductor-ferromagnet heterostructures, a ferromagnetic layer could seriously suppress or even kill the superconductivity of superconducting layers depending on their thickness \cite{Hetero_Bourgeois_2002, Hetero_Sidorenko_2003, Hetero_Gong_2015}. Another different case is a contact between a ferromagnet and a bulk superconductor, where the bulk superconductivity can be preserved and thus can be used to measure the spin polarization of the adjacent ferromagnet \cite{SpinPolar_Ji_2002}. However, those experiments are all quite different from our situation, in which interfacial superconductivity could be induced by the ferromagnetic tips even with similar $T_c$s as the non-magnetic tips. The compatibility of induced superconductivity and ferromagnetism suggests an unconventional pairing component \cite{PhysRevB.22.3173, PhysRevB.86.214514, PhysRevB.89.014506}, which deserves further experimental and theoretical investigations.


In order to get further insight into the observed TISC, the measured Andreev-reflection spectra were studied in detail as exemplified in Fig.~\ref{fig:BTK}(a). A revised two-channel BTK model was adopted \cite{15BTK, 16Dynes, 17BTK_Gamma, 18d-BTK} to take into account the contribution from both superconducting and normal channels, in which the parameters of $\Delta$, $Z$, $\Gamma$ and $W_S$ are superconducting gap, effective barrier height, broadening factor and the spectral weight contributed by the superconducting channel. Thus the total conductance $C_{tot}$ can be written as $C_{tot}=W_s\cdot C_{s}+(1-W_s)$ (having been normalized according to the normal conductance $C_n$). It was found that, the simulation with an appropriate $W_s=21\%$ can fit the data of Fig.~\ref{fig:BTK}(a) very well, whereas neither a single channel ($W_s=100\%$) or an underestimated value of $W_s$ (say, $10\%$) cannot. Moreover, as shown in Fig.~\ref{fig:BTK}(b), a slight adjustment of junction resistance could lead to a change of $W_s$ without altering the nature of the superconducting channel. These analyses indicate the validity of the multiple-channel model and the finite size of the superconducting area beneath the tip. Combining the resistance of a superconducting channel ($R_S$) and its effective barrier $Z$, we can evaluate the lateral dimension ($a$) of the superconducting channel using the well-known expression of Sharvin resistance for a micro-constriction in the ballistic limit, i.e., $R_S=(1+Z^2)\cdot \rho l/4a^2$ \cite{15BTK}, where $\rho$ is the bulk resistivity of the two banks of a PC. As a rough estimate, $\rho$ is taken as $0.2 \mu\Omega\cdot$cm by considering $\rho(WC)\sim0.4 \mu\Omega\cdot$cm \cite{23WC_MR} and $\rho(Au)\sim0.02 \mu\Omega\cdot$cm \cite{Resistivity_Gold_1979}, and the mean free path $l$ can be approximated using the above resistivity \cite{MFP_Nb3Sn_1979} to be $3.2 \mu$m and $3.0 \mu$m for WC and Au, respectively. From the fitting in Fig.~\ref{fig:BTK}(a), $R_S\approx21\Omega$ can be determined by $R_J/W_s$ with the total junction resistance of $R_J \approx 4.4 \Omega$. All the parameters lead to a lateral dimension of the superconducting channel as low as 10 nm. Considering that the contact size in thermal regime is much larger than the mean free path, we could give a conservative estimation that the induced local superconductivity could have a dimension varying from tens of nanometers to the order of micrometer. On the one hand, the magnetic compatible superconductivity achieved at nano-scale apparently contradicts the conventional Bardeen-Cooper-Schrieffer (BCS) theory with spin-singlet pairing \cite{Hetero_Bourgeois_2002, Hetero_Sidorenko_2003, Hetero_Gong_2015}. On the other hand, the superconducting region realized in micron degree will make it possible to be extended to a much larger scale by using appropriate ways such as metal film deposition.  Figure \ref{fig:BTK}(c) demonstrates the global fitting of a series of temperature dependent spectra. The superconducting gaps obtained from all such fittings are summarized in Fig.~\ref{fig:BTK}(d). The temperature dependencies of the superconducting gaps do not merge into each other but distribute randomly around the BCS law, which could be a consequence of multi-band superconductivity \cite{review_MultiBand} or the unconventional property of the TISC.

\begin{figure}[]
\includegraphics[scale=0.5]{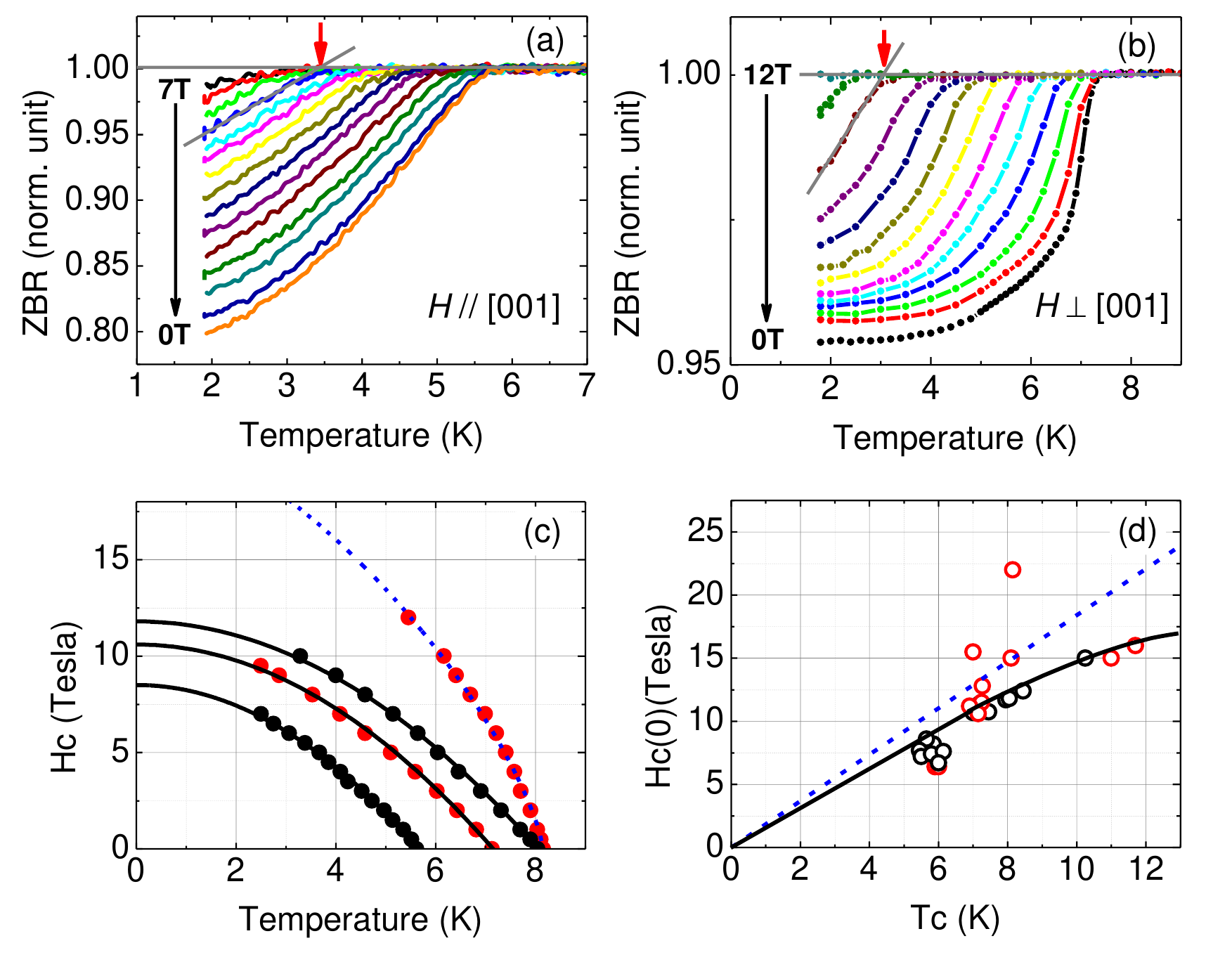}
\caption{\label{fig:PDs} (Color online)  (a) and (b) Typical temperature evolutions of zero-bias resistance taken from the PC spectra measured at various magnetic fields oriented along or perpendicular to the [001] axis of the sample, i.e., $H^{//}$ or $H^\perp$. The critical fields $H_c(T)$ can be determined by extrapolating the steepest part of the superconducting transition to intersect the horizontal normal-state background, as indicated by the arrows.  (c) Determined critical fields $H_c(T)$ of some PCs for either $H^{//}$ (black symbols) or $H^\perp$ (red symbols). The solid lines obey the empirical expression of $H_c(T)=H_c(0)[1-(T/T_c)^2]$, and the dashed line represents a trial formula of $H_c(T)=H_c(0)[1-(T/T_c)^{1.5}]^{0.75}$. (d) Summary of $H_c(0)$ versus $T_c$ obtained from various point contacts. Black and red symbols correspond to the configurations of $H^{//}$ and $H^\perp$, respectively. The dashed line indicates the Pauli paramagnetic limit and the solid line is a guide to the eye. }
\end{figure}

The TISC on WC in addition to Cd$_3$As$_2$ \cite{1Cd3As2_GS, 2Cd3As2_WJ} and TaAs \cite{3TaAs_GS, 4TaAs_WJ} implies a common mechanism shared by these topological semimetals. Among several possible explanations, the first candidate is band structure modifications induced by tip pressure as proposed in previous studies. However, this speculation is inconsistent with our finding that a soft Au tip could induce a $T_c$ of 11.5 K on WC, even higher than that of 7.5 K achieved by a W tip. In fact, WC has a hardness of near 2000 kg/mm$^2$ (along c-axis) at room temperature, which is much higher than that of Au (100-200 kg/mm$^2$). Moreover, no superconductivity above 1.5 K can be observed in our WC samples under a very high hydrostatic pressure up to 11 GPa. Therefore, tip pressure effect can be ruled out in our experiments. Meanwhile, as aforementioned, the lateral size of the superconducting puddles can be as large as the order of micrometer, implying the confinement effect is also irrelevant here. So, the TISC observed here is closely related to the interface coupling between normal metals and WC. In this case, a two dimensional superconductivity might be involved \cite{review_2D-SC}. To examine the expected 2D scenario, we have measured critical magnetic field for different configurations of $H//[001]$ and $H\perp[001]$ (henceforth referred to as $H^{//}$ and $H^{\perp}$ ). As exemplified by Figs.~\ref{fig:PDs}(a)-(c), the determined $H_c(T)$ relations obviously deviate from the 2D regime dictating $H^{//}_c(T)\propto1-T/T_c$ and $H^{\perp}_c(T)\propto(1-T/T_c)^{1/2}$. It was found that all $H^{//}_c(T)$s and most $H^{\perp}_c(T)$s can be approximated by the empirical formula of $H_c(T)=H_c(0)[1-(T/T_c)^2]$. Whereas in some cases, $H^{\perp}_c(T)$ can be described by a trial formula of $H_c(T)=H_c(0)[1-(T/T_c)^{1.5}]^{0.75}$ which lies between the empirical square law and the 2D square root law. This suggests that the induced superconductivity locates at the boundary of the 3D and 2D regimes, which might be ascribed to a superconducting layer with a significant thickness. All $H_c(0)$ determined by extrapolating $H_c(T)$ to zero temperature are plotted against $T_c$ in Fig.~\ref{fig:PDs}(d). Although most data follow a common trend indicated by the black solid line, some points of $H^{\perp}_c(0)$ have much larger values exceeding Pauli paramagnetic limit. 


In summary, we have observed TISC up to near 12 K on the newfound topological semimetal WC with super hardness. The estimated lateral dimensions of the superconducting regions vary from dozens of nanometers to the order of micrometer, and the determined temperature dependencies of superconducting gap and critical field behave differently from the conventional superconductivity. Most surprisingly, the achieved critical temperature is insensitive to the ferromagnetism of the tip materials. These findings not only reveal the unconventional nature of the TISC but also support interface coupling instead of tip-pressure or confinement effect as the dominant origin. Furthermore, the high chemical stability and super hardness of WC offer an opportunity to fabricate metal/WC interface and explore exotic quantum phenomena in this topological material.


\begin{acknowledgments}
This work was financially supported by the National Key R\&D Program of China (2017YFA0302904, 2018YFA0305602, 2016YFA0300604, 2015CB921303, 2018YFA0305700, 2017YFA0303201), the National Natural Science Foundation of China (11574372, 11322432, 11404175, 11874417, 11804379, 11674331), the National Basic Research Program of China 973 Program (Grant No.2015CB921303), the ``Strategic Priority Research Program (B)" of the Chinese Academy of Sciences ( Grant No. XDB07020300, XDB07020100), the¡®100 Talents Project¡¯of the Chinese Academy of Sciences; CASHIPS Director's Fund (No. BJPY2019B03) and The Recruitment Program for Leading Talent Team of Anhui Province (2019-16).
\end{acknowledgments}

\bibliography{WC_rev}

\end{document}